\documentclass[aps, twocolumn, showpacs, amsmath,superscriptaddress]{revtex4}
\usepackage{dcolumn}
\usepackage{bm}
\usepackage{graphicx}
\usepackage{color}
\usepackage{ulem}
\DeclareGraphicsExtensions{. jpg,. pdf, . mps, . png, . eps, . ps, . EPS}

\begin{document}
\def\be{\begin{equation}}
\def\ee{\end{equation}}

\def\bc{\begin{center}} 
\def\ec{\end{center}}
\def\bea{\begin{eqnarray}}
\def\eea{\end{eqnarray}}
\newcommand{\avg}[1]{\langle{#1}\rangle}
\newcommand{\ket}[1]{\left |{#1}\right \rangle}
\newcommand{\beq}{\begin{equation}}
\newcommand{\eneq}{\end{equation}}
\newcommand{\beqnn}{\begin{equation*}}
\newcommand{\eneqnn}{\end{equation*}}
\newcommand{\beqy}{\begin{eqnarray}}
\newcommand{\eneqy}{\end{eqnarray}}
\newcommand{\beqynn}{\begin{eqnarray*}}
\newcommand{\eneqynn}{\end{eqnarray*}}
\newcommand{\half}{\mbox{$\textstyle \frac{1}{2}$}}
\newcommand{\proj}[1]{\ket{#1}\bra{#1}}
\newcommand{\av}[1]{\langle #1\rangle}
\newcommand{\braket}[2]{\langle #1 | #2\rangle}
\newcommand{\bra}[1]{\langle #1 | }
\newcommand{\Avg}[1]{\left\langle{#1}\right\rangle}
\newcommand{\inprod}[2]{\braket{#1}{#2}}
\newcommand{\upket}{\ket{\uparrow}}
\newcommand{\downket}{\ket{\downarrow}}
\newcommand{\Tr}{\mathrm{Tr}}
\newcommand{\hcontrol}{*!<0em,.025em>-=-{\Diamond}}
\newcommand{\hctrl}[1]{\hcontrol \qwx[#1] \qw}
\newenvironment{proof}[1][Proof]{\noindent\textbf{#1.} }{\ \rule{0.5em}{0.5em}}
\newtheorem{mytheorem}{Theorem}
\newtheorem{mylemma}{Lemma}
\newtheorem{mycorollary}{Corollary}
\newtheorem{myproposition}{Proposition}
\newcommand{\vp}{\vec{p}}
\newcommand{\Or}{\mathcal{O}}
\newcommand{\so}[1]{{\ignore{#1}}}

\newcommand{\red}[1]{\textcolor{red}{#1}}
\newcommand{\blue}[1]{\textcolor{blue}{#1}}

\title{Emergence of overlap in ensembles of spatial multiplexes \\and statistical mechanics of spatial interacting network ensembles}

\author{Arda Halu} 
\affiliation{Department of Physics, Northeastern University, Boston, MA,USA} 
\author{Satyam  Mukherjee}
\affiliation{Kellogg School of Management, Northwestern University, Evanston IL, USA}
\author{Ginestra Bianconi}

\affiliation{School of Mathematical Sciences, Queen Mary University of London, London E1 4NS, UK}

\begin{abstract}
Spatial networks range from the  brain networks, to transportation networks and infrastructures.  Recently interacting and multiplex networks are attracting great attention because their dynamics and robustness cannot be understood without treating at the same time several networks. Here we present  maximal entropy ensembles  of  spatial multiplex and spatial interacting networks that can be used in order to model spatial multilayer network structures and to build null models of real datasets.
We show that spatial multiplexes naturally develop a significant overlap of the links, a noticeable property  of many multiplexes that can affect significantly the dynamics taking place on them. Additionally, we characterize ensembles of  spatial interacting networks and we analyze  the structure of interacting airport and railway networks in India, showing the effect of space in determining the link probability.
\end{abstract}

\pacs{89.75.Hc,89.75.-k,89.75.Fb}

\maketitle
\section{Introduction}
Many real networks \cite{Barthelemy_rev}  are embedded in a real \cite{Bullmore,railway,Colizza,Vito_road,Manna} or in a hidden space \cite{Arenas, Hyperbolic} which plays a key role in determining their topology. Major examples of spatial networks are  brain networks \cite{Bullmore},  infrastructures \cite{railway,Colizza}, road networks \cite{Vito_road}, and social networks \cite{Arenas}. In many of these cases the networks are also multiplex indicating that the $N$ nodes of the system can be connected by links of different nature forming a multilayer structure of networks.
For example, two cities can be linked at the same time by a train connection and  flight connection, or in social networks people can be linked at the same time by friendship relation, scientific collaborations etc. In physiology, the  brain network interacts with the circulatory system that provides the blood supply to the  brain.
The field of multiplex networks is attracting recent attention. New multiplex datasets  \cite{Garlaschelli,Mucha, Thurner,Kurths,Boccaletti_air,Vito_m} and multiplex network measures have been introduced in order to quantify their complexity. Examples of  such measures are  the overlap \cite{Thurner,Boccaletti_air,Vito_m} of the links in different layers, the interdependence \cite{Marc,Kurths} that extends the concept of betweenness centrality to multiplexes, or the centrality measures \cite{centrality,Pagerank}. Many dynamical processes have been defined on multiplexes, including cascades of failure in interdependent networks \cite{Havlin1,Havlin2,Havlin3,Son,goh2}, antagonistic percolation \cite{JSTAT}, dynamical cascades \cite{Leicht}, diffusion \cite{Diffusion}, epidemic spreading \cite{Boguna_spr}, election models \cite{Elections},  game theory \cite{Cooperation,Perc},  etc.
Moreover multiplex network models are starting to be proposed following equilibrium or non-equilibrium approaches \cite{PRE,PRL,Goh,tensor}. In this context it has been found \cite{PRE} that the extension of the configuration model to uncorrelated multiplex contains a vanishing overlap in the thermodynamic limit.

Building on the statistical mechanics of network ensembles \cite{Newman1,Newman2,entropy,Kartik1,PNAS,Peixoto1,Peixoto2}, here  we characterize the statistical mechanics  of spatial multiplex ensembles. These ensembles of multiplexes can be used for generating multiplexes with given structural properties or for randomizing given spatial multiplex datasets and have potential impact modelling and inference of spatial multiplexes. Here we show a noticeable property of spatial multiplexes: these multilayer structures  in which  the nodes are positioned in a real or in a hidden space, naturally allow for the emergence of the overlap.
This phenomenon can explain why a significant overlap is observed so often in multiplex datasets \cite{Thurner,Boccaletti_air,Vito_m} and  might have different implications for brain networks, transportation networks, social networks and in general any spatial multiplex. In fact it has been observed that the outcome of the dynamical processes depends significantly  on the presence of the overlap \cite{cellai,havlin_o}.

Moreover we characterize  interacting networks ensembles in which the networks in the complex multilayer structure have a different set of nodes, and we apply this approach to characterize the airport network \cite{bagler} and the railway network in India updating in this way the analysis of the railway network in India performed ten years ago \cite{railway}.
We observe that the airport network and the railway networks have different degree distributions and different degree correlations. Nevertheless the function $W(d)$ modulating the link probability with the distance between the nodes, decays as a power-law with distance for large distances, i.e. $W(d)\propto d^{-\delta}$. This indicates that in both networks long distance connections are significantly represented improving the navigability of the two interacting networks. Moreover it suggests that these networks can be considered as maximal entropy networks associated with a given cost of the connections depending logarithmically with the distance between the linked nodes.

The paper is structured as follows: in section II we review the general derivation  of spatial network ensembles, and we give major specific examples; in section III  we present multiplex ensembles and we define the total and local overlap between two layers showing  that uncorrelated multiplex ensemble have a negligible overlap; in section IV we present  spatial multiplex ensembles and we show that these multiplex naturally  develop a significant overlap of the links, providing one major example and leaving to the appendix the characterization of other examples; in section V we define interacting networks,  we present a derivation of interacting network ensembles, and  we characterize a new dataset of interacting air and train transportation networks in India. Finally in section VI we give the conclusions.
\section{ Spatial network ensembles }
\subsection{General derivation}
An important framework to  model  complex networks is the one of network ensembles \cite{Newman1,Newman2,entropy,BC,Kartik1,PNAS,Peixoto1,Peixoto2}.
In this context we model an {\it ensemble} of networks with given structural properties by giving a probability $P(G)$ to each network $G=(V,E)$ of the ensemble. For this ensemble the {\it entropy} $S$ quantifies the logarithm of the typical number of networks represented in the ensemble and is given by 
\begin{equation}
S=-\sum_{G} P(G)\log P(G).
\label{S}
\end{equation}
The entropy also quantify the {\it complexity} of the ensemble taken into consideration.
Suppose that we want to construct a network ensemble satisfying a set $K$ of soft constraints (constraints satisfied in average) 
\begin{equation}
\sum_{G}F_{\mu}(G)P(G)=C_{\mu},
\label{constr}
\end{equation}
with $\mu=1,2\ldots, K$, and $F_{\mu}(G)$ being a function of the network. For example $F_{\mu}(G)$ can be the total number of links or the degree of a node of the network. The least biased way of constructing a network ensemble satisfying these constraints is by maximizing the entropy $S$ given by Eq. $(\ref{S})$ under the constraints given by Eqs. $(\ref{constr})$. 
By introducing the Lagrangian multipliers $\lambda_{\mu}$ and maximizing the entropy, we get that the probability for a network in this network  ensemble is given by the exponential
\bea
P(G)=\frac{1}{Z}e^{-\sum_{\mu=1}^K \lambda_{\mu}F_{\mu}(G)},
\eea
where $Z$ is the normalization constant, and the values of the Lagrangian multipliers $\lambda_{\mu}$ for each constraint $\mu=1,2,\ldots, K$ are fixed by imposing the constraints in Eqs. $(\ref{constr})$.
We note here that this specific type of ensemble is also called exponential random network ensemble (due to the exponential expression of $P(G)$) or canonical network  ensemble (because the constraints $F_{\mu}(G)$ are only satisfied in average).
If we indicate by $a_{ij}$ the matrix element $(i,j)$ of the adjacency matrix of a generic network in the ensemble, in this ensemble the probability of a link between node $i$ and node $j$ is given by 
\bea
p_{ij}=\avg{a_{ij}}=\sum_{G} a_{ij} \frac{1}{Z}e^{-\sum_{\mu=1}^K \lambda_{\mu}F_{\mu}(G)}.
\eea

Let us now consider  spatial network ensembles where the node of the network are embedded in a geometric space.
To this end, we assume that the nodes of the network are embedded in a geometrical space with each node $i=1,2,\ldots, N$ positioned at a point of coordinates $\vec{r}_i$. Therefore  we can define for each pair of nodes $i$ and $j$ a distance $d_{ij}$.
The probability of a network in the spatial ensemble is conditioned on the values of the coordinates of the nodes, i.e. strictly speaking we have a $P(G|\{\vec{r}_i\})$ where $\{\vec{r}_i\}$ is the complex set of the coordinates of the nodes in the geometrical embedding space.
For ensembles of spatial networks the  {\it entropy} $S$  is given by 
\begin{equation}
S=-\sum_{G} P(G|\{\vec{r}_i\})\log P(G|\{\vec{r}_i\}).
\label{S2}
\end{equation}
Spatial network ensembles can be constructed by maximizing the 
entropy of the ensemble, while fixing a set  $K$ of soft constraints 
\begin{equation}
\sum_{G}F_{\mu}(G|\{\vec{r}_i\})P(G|\{\vec{r}_i\})=C_{\mu},
\label{constr2}
\end{equation}
with $\mu=1,2\ldots, K$, where $F_{\mu}(G|\{\vec{r}_i\})$ is a function of the network and the positions of the nodes.
In this way it is easy to show that the probability $P(G|\{\vec{r}_i\})$ of a network in this ensembles is given by 
\bea
P(G|\{\vec{r}_i\})=\frac{1}{Z}e^{-\sum_{\mu=1}^K \lambda_{\mu}F_{\mu}(G|\{\vec{r}_i\})},
\label{pg2}
\eea
where $Z$ is the normalization constant, and the values of the Lagrangian multipliers $\lambda_{\mu}$ for each constraint $\mu=1,2,\ldots, K$ are fixed by imposing the constraints in Eqs.~$(\ref{constr2})$.
\subsection{Specific examples}
\subsubsection{Spatial network ensembles with fixed  expected number of links at a given distance}
Maximal entropy network ensembles or exponential random networks are not only interesting in order to model a certain class of networks, but provide also a well defined framework to construct null network models starting from a real network realization \cite{entropy}.
In this context we can call these ensembles also randomized networks ensembles.  
Let us assume, for example, to have a given undirected spatial network, and to desire to construct randomized versions of it satisfying a set of constraints: the way to do this is by sampling the maximum entropy ensemble. 
In the construction of a randomized version of a spatial network, in many occasions  it is interesting to consider  networks satisfying at the same time the following constraints:
\begin{itemize}
\item {\it (a)}
 the expected degree sequence in the network ensemble is equal to the degree sequence of the given network;
 \item {\it (b)} the number of expected links connecting nodes at a given distance is equal to the number of such links observed in the given network.
 \end{itemize}
In this case the set of constraints $F_{\mu}(G|\{\vec{r}_i\})$ are given by the following conditions.
\begin{itemize}
\item {\it (a)} The conditions on the expected average degrees can be expressed as
  \bea
\kappa_i&=&\sum_{G}P(G|\{\vec{r}_i\})F_{i}(G)\nonumber \\&=&\sum_{G}P(G|\{\vec{r}_i\})\sum_{j=1}^N a_{ij},
\label{k2}
\eea
for $\mu=i=1,2\ldots, N$ (where $\kappa_i$ is the expected degree of node $i$ in the ensemble).
\item{\it (b)} The conditions on the expected number of nodes at a given distance can be expressed as

\bea
n(d_{\mu})=&=&\sum_{G}P(G|\{\vec{r}_i\})F_{\mu}(G|\{\vec{r}_i\})\nonumber \\&=&\sum_{G}P(G|\{\vec{r}_i\})\sum_{i<j} a_{ij}\chi_{\mu}(d_{\mu},d_{ij}),
\label{nd}
\eea
where we have discretized the possible range of distances in bins  $(d_{\mu},d_{\mu}+\Delta_{\mu} d)$ with $\mu=N+1\ldots, K$. Here,  $\Delta_{\mu}d$ indicates the size of the $\mu$ bin (for example we can take bins of size increasing as a power-law of the distance $d_{\mu}$). Moreover in the Eq. $(\ref{nd})$, we have $\chi_{\mu}(d_{\mu},d_{ij})=1$ if $d_{ij}\in (d_{\mu},d_{\mu}+\Delta_{\mu} d)$ and    $\chi_{\mu}(d_{\mu},d_{ij})=0$ otherwise.
\end{itemize}
In this spatial network ensemble the probability $P(G|\{\vec{r}_i\})$ given by Eq. $(\ref{pg2})$ takes the simple form
\bea
P(G|\{\vec{r}_i\})=\prod_{ij}[p_{ij}(d_{ij})]^{a_{ij}}[1-p_{ij}(d_{ij})]^{1-a_{ij}}.
\eea
with 
\bea
p_{ij}(d_{ij})=
\frac{e^{-\lambda_i-\lambda_j-\sum_{\mu=N+1,K}\lambda_{\mu}\chi(d_{\mu},d_{ij})}}{1+e^{-\lambda_i-\lambda_j-\sum_{\mu=N+1,K}\lambda_{\mu}\chi_{\mu}(d_{\mu},d_{ij})}},
\label{pijd2}\eea
where the Lagrangian multipliers $\lambda_{\mu}$ are fixed by the conditions Eqs.~$(\ref{k2})-(\ref{nd})$.
Another way to write the link  probability in Eq. $(\ref{pijd2})$ is by putting $e^{-\lambda_i}=\theta_i$ and $e^{-\sum_{\mu=N+1,K}\lambda_{\mu}\chi(d_{\mu},d_{ij})}=W(d_{ij})$ and write 
\bea
p_{ij}(d_{ij})=\frac{\theta_i \theta_j W(d_{ij})}{1+\theta_i \theta_j W(d_{ij})}.
\label{pW}
\eea
In \cite{PNAS} the top 500 USA airport network \cite{Colizza} was considered and and the function $W(d)$ measured from the data. Interestingly enough, this function decays as a power-law of the distance for large distances, i.e. $W(d)\propto  d^{-\delta}$  with $\delta\simeq 3$ \cite{PNAS}.

\subsubsection{Spatial network ensemble with fixed  expected total cost of the links}
Many spatial networks, from brain networks to transportation networks have a cost associated to each link that is usually a function of the distance between the connected nodes.
Therefore here we consider network ensembles in which we fix the expected degree $\kappa_i$ for each node $i=1,2\ldots, N$ of the network
\bea
\kappa_i&=&\sum_{G}P(G|\{\vec{r}_i\})F_{i}(G)\nonumber \\&=&\sum_{G}P(G|\{\vec{r}_i\})\sum_{j=1}^N a_{ij},
\eea
and at the same time we fix a total cost  $L$ of the links. In particular $L$ is the sum of all the costs of the links $(i,j)$, $f(d_{ij})$ where we assume that these costs are  a function of the distance $d_{ij}$ between nodes.
Therefore  we take
\bea
L&=&\sum_G P(G|\{\vec{r}_i\}) F_{N+1}(G|\{\vec{r}_i\})\nonumber \\
&=&\sum_G P(G|\{\vec{r}_i\}) \sum_{i<j} f(d_{ij})a_{ij}.
\label{L}
\eea
In this spatial network ensemble the probability $P(G|\{\vec{r}_i\})$ given by Eq. $(\ref{pg2})$ takes the simple form
\bea
P(G|\{\vec{r}_i\})=\prod_{ij}[p_{ij}(d_{ij})]^{a_{ij}}[1-p_{ij}(d_{ij})]^{1-a_{ij}}.
\label{uno}
\eea
with 
\bea
p_{ij}(d_{ij})=\frac{e^{-\lambda_i-\lambda_j-\lambda_{N+1} f(d_{ij})}}{1+e^{-\lambda_i-\lambda_j-\lambda_{N+1} f(d_{ij})}}.
\eea
The function $f(d)$ can be chosen arbitrarily. Nevertheless typical functions that can be considered include the distance, and the logarithm of the distance, i.e.
\begin{eqnarray}
f(d_{ij})&=&d_{ij} \label{c1}\\
f(d_{ij})&=&\log d_{ij}.\label{c2}
\end{eqnarray}
These two expressions lead respectively to the following probability of the link between node $i$ and node $j$.
\begin{eqnarray}
p_{ij}(d_{ij})&=&\frac{e^{-\lambda_i-\lambda_j-d_{ij}/d_0}}{1+e^{-\lambda_i-\lambda_j-d_{ij}/d_0}}\label{pex} \\
p_{ij}(d_{ij})&=&\frac{e^{-\lambda_i-\lambda_j}d_{ij}^{-\delta}}{1+e^{-\lambda_i-\lambda_j}d_{ij}^{-\delta}}\label{ppow}
\end{eqnarray}
where the $N+1$ Lagrangian multiplier enforcing the constraint Eq.$(\ref{L})$ is given by $\lambda_{N+1}=1/d_0$ in the first case and $\lambda_{N+1}=\delta$ in the second case.
The Lagrangian multipliers $\lambda_i$ with $=1,2\ldots, N$ enforce the conditions over the expected degree of the node $i$.
The probabilities Eq. $(\ref{pex})$ and Eq. $(\ref{ppow})$ and be also be expressed in terms of $\theta_i=e^{-\lambda_i}$, i.e.
\bea
p_{ij}(d_{ij})&=&\frac{\theta_i \theta_j e^{-d_{ij}/d_0}}{1+\theta_i \theta_j e^{-d_{ij}/d_0}}\label{pex2} \\
p_{ij}(d_{ij})&=&\frac{\theta_i \theta_j d_{ij}^{-\delta}}{1+\theta_i\theta_j d_{ij}^{-\delta}}\label{ppow2}
\eea
where $(\{\theta_i\},d_0)$ or $(\{\theta_i\},\delta)$ are also called ``hidden variables".
Therefore if we analyse a real network dataset considering the randomized network ensemble with expected number of links at a given distance (as we have done in the previous subsection) and we observe a probability distribution given by Eq. $(\ref{pW})$ with $W(d)\propto d^{-\delta}$ we can deduce that the network can be thought as maximal entropy network with an associated cost of the links given by Eq.$(\ref{L}),(\ref{c2})$, while if we observe $W(d)\propto e^{-d/d_0}$ the network ensemble can be thought as a maximal entropy network ensembles with an associated cost of the links given by Eqs. $(\ref{L}),(\ref{c1})$.

\subsubsection{Spatial bipartite network ensemble with fixed expected number of links at a given distance}
Spatial networks can be of different types: directed, weighted, with features of the nodes, etc.
An interesting case that we will consider here is the case in which the spatial network is bipartite.
In particular, in this subsection we will  define  maximal entropy ensembles of bipartite spatial networks.
Let us suppose that $b_{ij}$ is the incidence matrix of the bipartite network, with $i=1,2,\ldots, N_1$ and $j=1,2\ldots N_2$ indicating distinct nodes of coordinates $\{\vec{r}^1_i\}$ and $\{\vec{r}^2_j\}$ respectively.

As an example of a bipartite spatial network ensemble we consider the network in which we fix the expected degree $\{\kappa^1_i\}$ of nodes $i=1,2\ldots N_1$ and the expected degree $\{\kappa_j^2\}$ of nodes $j=1,2\ldots, N_2$ and in addition to this we fix the expected number of links at a given distance.
In particular the soft constraints that we impose on the ensemble are given by the following list.
\begin{itemize}
\item {\it (a)} The conditions on the expected average degrees $\{\kappa_i^1\} $can be expressed as
  \bea
\kappa_i^1&=&\sum_{G}P(G|\{\vec{r}_i^1\},\{\vec{r}_j^2\})\sum_{j=1}^{N_2} b_{ij},
\label{k11}
\eea
for $i=1,2\ldots, N_1$. These are the conditions $\mu=1,2,\ldots, N_1$.
\item {\it (b)} The conditions on the expected average degrees $\{\kappa_j^2\} $can be expressed as
  \bea
\kappa_j^2&=&\sum_{G}P(G|\{\vec{r}_i^1\},\{\vec{r}_j^2\})\sum_{i=1}^{N_1} b_{ij},
\label{k22}
\eea
for $j=1,2\ldots, N_2$. These are the conditions $\mu=N_1+1,\ldots, N_1+N_2$.
\item{\it (c)} The conditions on the expected number of nodes at a given distance can be expressed as
\bea
\hspace*{-5mm} n(d_{\mu})&=&\sum_{G}P(G|\{\vec{r}_i^1\},\{\vec{r}_j^2\})\sum_{i=1}^{N_1}\sum_{j=1}^{N_2} b_{ij}\chi(d_{\mu},d_{ij}),
\label{nd2}
\eea
where we have discretized the possible range of distances in bins  $(d_{\mu},d_{\mu}+\Delta_{\mu} d)$ with $\mu=N_1+N_2+1,\ldots, K$. Moreover in the Eq. $(\ref{nd2})$, we have $\chi_{\mu}(d_{\mu},d_{ij})=1$ if $d_{ij}\in (d_{\mu},d_{\mu}+\Delta_{\mu} d)$ and    $\chi_{\mu}(d_{\mu},d_{ij})=0$ otherwise.
\end{itemize}
Following the same type of approach described by the previous cases, we can show that 
\bea
P(G|\{\vec{r}_i^1\},\{\vec{r}^2_j\})=\prod_{ij}[p_{ij}(d_{ij})]^{b_{ij}}[1-p_{ij}(d_{ij})]^{1-b_{ij}}.
\eea
with 
\bea
p_{ij}(d_{ij})=
\frac{e^{-\lambda_i-\lambda_{N_1+j}-\sum_{\mu=N_1,N_2+1,K}\lambda_{\mu}\chi(d_{\mu},d_{ij})}}{1+e^{-\lambda_i-\lambda_{N_1+j}-\sum_{\mu=N_1+N_2+1,K}\lambda_{\mu}\chi(d_{\mu},d_{ij})}},
\label{pijd22}\eea
where the Lagrangian multipliers $\lambda_{\mu}$ are fixed by the conditions Eqs.~$(\ref{k11})-(\ref{k22})-(\ref{nd2})$.
Another way to write the link  probability in Eq. $(\ref{pijd22})$ is by putting $e^{-\lambda_i}=\theta^1_i$ $e^{-\lambda_{N_1+j}}=\theta^2_j$ and $e^{-\sum_{\mu=N_1+N_2+1,K}\lambda_{\mu}\chi(d_{\mu},d_{ij})}=W(d_{ij})$ and write 
\bea
p_{ij}(d_{ij})=\frac{\theta^1_i \theta^2_j W(d_{ij})}{1+\theta^1_i \theta^2_j W(d_{ij})}.
\label{pWb}
\eea

\section{  Multiplexes}
\subsection{Definition and overlap}
A multiplex is a multilayer structure formed by $M$ layers and $N$ nodes $i=1,2,\ldots, N$. Every node is represented in every layer of the multiplex. Every layer $\alpha=1,2,\ldots, M$ is formed by a network $G_{\alpha}=(V,E_{\alpha})$ with adjacency matrix of elements $a_{ij}^{\alpha}=1$ if there is a link between node $i$ and node $j$ in layer $\alpha$ and otherwise $a_{ij}^{\alpha}=0$.
 Here we introduce the definition of global and local overlap of the links, one of the major structural characteristics of a multiplex observed in several datasets \cite{Thurner,Boccaletti_air,Vito_m}

For two layers $\alpha,\alpha'$ of the multiplex   the {\it global overlap} $O^{\alpha,\alpha'}$ is defined as the total number of pairs of nodes connected at the same time by a link in layer $\alpha$ and a link in layer $\alpha'$, i.e. 
\begin{equation}
O^{\alpha,\alpha'}=\sum_{i<j} a_{ij}^{\alpha}a_{ij}^{\alpha'}.
\label{Og}
\end{equation}
Furthermore, for a node $i$ of the multiplex, the {\it local overlap} $o_i^{\alpha,\alpha'}$ of the links in two layers $\alpha$ and $\alpha'$ is defined as the total number of nodes $j$ linked to the node $i$ at the same time by a link in layer $\alpha$ and a link in layer $\alpha'$, i.e.
\begin{equation}
o_i^{\alpha,\alpha'}=\sum_{j=1}^N a_{ij}^{\alpha}a_{ij}^{\alpha'}.
\label{oi}
\end{equation}
In spatial networks we expect the global and local overlap to be significant. For example in transportation networks  within the same country, if we consider train and long-distance bus transportation we expect to observe a significant overlap. Also in case of social multiplex networks where each layer represents  different means of communication between people, (emails, mobile, sms, etc.) two people that are linked in one layer are also likely to be linked in another layer, forming a multiplex with significant overlap.
This observation is supported by the analysis of real multiplex datasets \cite{Thurner, Boccaletti_air,Vito_m}  that are characterized by a significant overlap of the links.
\subsection{Multiplex ensembles}

Recently, the research on multiplexes has been gaining large momentum. Different models for capturing the structure of multiplexes have been proposed, including multiplex ensembles \cite{PRE}, growing multiplex models \cite{PRL,Goh} and models based on tensor formalism \cite{tensor}.

Multiplex ensembles describe maximal entropy multiplexes satisfying specific structural constraints, and are proposed to be very efficient null models for describing real multiplexes with different features.
A multiplex ensemble is determined once the probability $P(\vec{G})$ of the multiplex $\vec{G}=(G^1,G^2,\ldots,G^{\alpha},\ldots, G^M)$ is fixed.
The entropy of the multiplex ensemble $S$ is given by 
\bea
S=-\sum_{\{\vec{G}\}}P(\vec{G})\log P(\vec{G})
\label{sm}
\eea
 and the maximum entropy multiplex ensembles can be defined as a function of the soft constraints we plan to impose on the ensemble \cite{PRE}.
We assume to have $K$ of such constraints determined by the conditions
\begin{equation}
\sum_{\vec{G}}P(\vec{G})F_{\mu}(\vec{G})=C_{\mu}
\label{constraints}
\end{equation}
with $\mu=1,2\ldots, K$, and  $F_{\mu}(\vec{G})$ determining the  structural constraints that we want to impose on the multiplex. For example, $F_{\mu}(\vec{G})$ can be equal to  the total number of links in a layer of  the multiplex $\vec{G}$ or the degree of a node in a layer of the multiplex $\vec{G}$ ( for a detailed account see \cite{PRE}). 
Maximizing the entropy given by Eq. $(\ref{sm})$  while satisfying the constraints given by Eqs.~$(\ref{constraints})$ we find that the probability of a multiplex $P(\vec{G})$ in the  multiplex ensemble is given by 
\begin{equation}
P(\vec{G})=\frac{1}{Z}\exp\left[-\sum_{\mu}\lambda_{\mu}F_{\mu}(\vec{G})\right],
\label{PC}
\end{equation}
where $Z$ is the normalization constant, and the Lagrangian multipliers $\lambda_{\mu}$ are fixed by the constraints in Eqs. $(\ref{constraints})$.

\subsection{Uncorrelated multiplex ensembles and their overlap}
Uncorrelated multiplex ensembles have a probability $P(\vec{G})$ that can be factorized into the probability of single networks, i.e.
\bea
P(\vec{G})=\prod_{\alpha=1}^M P_{\alpha}(G_{\alpha}).
\label{unc}
\eea
These ensembles are maximal entropy multiplex ensembles in which every soft constraint involves just a single network.
Furthermore in many cases the constraints are linear in the adjacency matrix. Examples of such constraints are the cases in which we fix the expected degree sequence, or the number of nodes between communities.
In these cases the probability $P_{\alpha}(G_{\alpha})$ take the simple expression
\bea
P_{\alpha}(G_{\alpha})=\prod_{i<j}[p_{ij}^{\alpha}{a_{ij}}+(1-p_{ij}^{\alpha})({1-a_{ij}})].
\label{pga}
\eea
An important example of such multiplexes is the one in which we fix the expected degree $\kappa_i^{\alpha}$ of each node $i$ in each layer $\alpha$ and we impose the structural cutoff $\kappa_i^{\alpha}<\sqrt{\avg{\kappa^{\alpha}}N}$.
In this case we have 
\bea
p_{ij}^{\alpha}=\frac{\kappa_i^{\alpha} \kappa_j^{\alpha}}{\avg{\kappa^{\alpha}}N}.
\label{punc}
\eea
If the multiplex ensemble is uncorrelated and $P_{\alpha}(G_{\alpha})$ is given by Eq. $(\ref{pga})$, we can easily calculate the average global overlap $\avg{O^{\alpha,\alpha'}}$ between two layers $\alpha$ and $\alpha'$ and the average local overlap $\avg{o^{\alpha,\alpha'}_i}$ between two layers $\alpha$ and $\alpha'$ where the global overlap $O^{\alpha,\alpha'}$ is defined in Eq.~$(\ref{Og})$ and the local overlap $o^{\alpha,\alpha'}_i$ is defined in Eq.~$(\ref{oi})$. These quantities are given by 
 \begin{eqnarray}
 \avg{O^{\alpha,\alpha'}}&=&\sum_{i<j}p^{\alpha}_{ij} p^{\alpha'}_{ij}\nonumber \\
 \avg{o_i^{\alpha,\alpha'}}&=&\sum_{j=1,j\neq i}^Np^{\alpha}_{ij} p^{\alpha'}_{ij}.
\label{AO} 
\end{eqnarray}
 
 For multiplex ensembles with given expected degree of the nodes in each layer,  with $p_{ij}^{\alpha}$ given by Eq. $(\ref{punc})$  we have
 \begin{eqnarray}
 \avg{O^{\alpha,\alpha'}}&=&\frac{1}{2} \left(\frac{\Avg{\kappa^{\alpha}\kappa^{\alpha'}}^2}{\Avg{\kappa^{\alpha}}\Avg{\kappa^{\alpha'}}}\right)\nonumber \\
 \avg{o_i^{\alpha,\alpha'}}&=&\kappa_i^{\alpha}\kappa_i^{\alpha'}\frac{\Avg{\kappa^{\alpha}\kappa^{\alpha'}}}{\Avg{\kappa^{\alpha}}\Avg{\kappa^{\alpha'}}N}
 \end{eqnarray}
 where $\Avg{\kappa^{\alpha}\kappa^{\alpha'}}=\sum_{i=1}^N \kappa_i^{\alpha}\kappa_i^{\alpha'}/N$.
 
 If the expected degrees in the different layers are uncorrelated (i.e. $\Avg{\kappa^{\alpha}\kappa^{\alpha'}}=\Avg{\kappa^{\alpha}}\Avg{\kappa^{\alpha'}}$) then the global and local overlaps are given by 
 \begin{eqnarray}
  \avg{O^{\alpha,\alpha'}}&=&\frac{1}{2}\left({\Avg{\kappa^{\alpha}}\Avg{\kappa^{\alpha'}}}\right)\ll N\nonumber \\
  \avg{o_i^{\alpha,\alpha'}}&=&\frac{\kappa_i^{\alpha}\kappa_i^{\alpha'}}{N}\ll \min(\kappa_i^{\alpha},\kappa_i^{\alpha'})
 \end{eqnarray}
 Therefore  in this case the overlap is negligible.
 Degree correlation in between different layers can enhance the overlap, but as long as $\Avg{\kappa^{\alpha}\kappa^{\alpha'}}\ll N$ the average global $ \avg{O^{\alpha,\alpha'}}$ and the local $\avg{o_i^{\alpha,\alpha'}}$ overlap continue to remain negligible with respect to the total number of nodes  in the two  layers and the degrees of the node $i$ in the two layers.
 Similarly  the expected global overlap  and local overlap is negligible  in the multiplex ensemble in which we fix at the same time the average degree of each node in each layer and the average number of links in between nodes of different communities in each  layer.
In general, as long as we have an uncorrelated multiplex with $P_{\alpha}(G_{\alpha})$ given by Eq. $(\ref{pga})$  and $p_{ij}^{\alpha}\ll 1, \forall (i,j)$, then the expected local and global overlap is negligible.
 The way to solve this problem is to consider correlated multiplexes. On one side it  is possible to model multiplexes with given set of multilinks, as described in \cite{PRE}, on the other side it is possible to consider spatial multiplexes as we will show in the next sections.
\section{Spatial multiplex ensembles}
\subsection{General derivation}
Spatial multiplexes are ensemble of networks $\vec{G}=(G_1,G_2,\ldots, G_M)$ where $M$ are the number of layers in the multiplex. Each network $G_{\alpha}=(V,E_{\alpha})$ with $\alpha=1,2\ldots, M$ is formed by the same $N$ nodes $i=1,2\ldots, N$ embedded in a metric space. Each node $i$ is assigned a coordinate $\vec{r}_i$ in this metric space.
 A spatial multiplex ensemble is defined once we define the probability $P(\vec{G}|\{\vec{r}_i\})$ of the multiplex $\vec{G}$ conditioned to the positions of the nodes $\{\vec{r}_i\}$.
For ensembles of spatial multiplexes the  {\it entropy} $S$  is given by 
\begin{equation}
S=-\sum_{\vec{G}} P(\vec{G}|\{\vec{r}_i\})\log P(\vec{G}|\{\vec{r}_i\}).
\label{S3}
\end{equation}
Spatial multiplex ensembles can be constructed by maximizing the 
entropy of the ensemble, while fixing a set  $K$ of soft constraints 
\begin{equation}
\sum_{\vec{G}}F_{\mu}(\vec{G}|\{\vec{r}_i\})P(\vec{G}|\{\vec{r}_i\})=C_{\mu},
\label{constr3}
\end{equation}
with $\mu=1,2\ldots, K$, and $F_{\mu}(\vec{G}|\{\vec{r}_i\})$ a function of the multiplex and the positions of the nodes.
In this way it is easy to show that the probability $P(\vec{G}|\{\vec{r}_i\})$ of a multiplex in this ensemble is given by 
\bea
P(\vec{G}|\{\vec{r}_i\})=\frac{1}{Z}e^{-\sum_{\mu=1}^K \lambda_{\mu}F_{\mu}(\vec{G}|\{\vec{r}_i\})},
\label{pg3}
\eea
where $Z$ is the normalization constant, and the values of the Lagrangian multipliers $\lambda_{\mu}$ for each constraint $\mu=1,2,\ldots, K$ are fixed by imposing the constraints in Eqs. $(\ref{constr3})$.
 A particular case of a spatial multiplex ensemble is generated by this approach when each constraint $F_{\mu}(\vec{G}|\{\vec{r}_i\})$ involves a single network in one layer of the multiplex. In this case $P(\vec{G}|\{\vec{r}_i\})$ can be written as 
 \bea
 P(\vec{G}|\{\vec{r}_i\})=\prod_{\alpha=1}^M P_{\alpha}(G_{\alpha}|\{\vec{r}_i\}).
 \label{su}
 \eea
 In this case the multiplex is {\it not}  uncorrelated because the probabilities $P_{\alpha}(G_{\alpha}|\{\vec{r}_i\})$ appearing in Eq. $(\ref{su})$ are conditioned on the position of the nodes $\{\vec{r}_i\}$ that are the same for every network $\alpha$.
 In particular,  unlike in the case in which we have Eq.~$(\ref{unc})$, these types of spatial multiplex might show a significant overlap of the links as we will show in the next subsections.

 \subsection{Expected overlap of spatial multiplexes}
 Many spatial multiplexes naturally develop  a significant overlap. Let us consider for simplicity  spatial multiplex ensembles in which every given multiplex has a probability  given by Eq.~$(\ref{su})$ where the probabilities $P_{\alpha}(G_{\alpha}|\{\vec{r}_i\})$ are given by Eq. $(\ref{pg2})$. The goal of this section is to show that these multiplexes, unlike uncorrelated multiplexes satisfying Eq. $(\ref{unc})$ can have a significant overlap. In the following subsection will focus our attention on multiplex ensembles with link probability decaying exponentially with distance and we will refer the interested reader to the appendix for the generalization of this derivation to multiplex with links decaying as a power-law of the distance or with different layers characterized by different spatial behavior (some layers with link probability decaying exponentially with distance and some layers with links probability decaying as a power-law of the distance).
 \subsubsection{Multiplex ensembles with link probability decaying exponentially with distance}
 \label{par}
In this subsection  we evaluate the expected overlap for  a multiplex where each $P_{\alpha}(G_{\alpha}|\{\vec{r}_i\})$ is given by 
 Eq.~$(\ref{uno})$ that we rewrite here for convenience,
 \bea
\hspace*{-5mm} P_{\alpha}(G_{\alpha}|\{\vec{r}_i\})=\prod_{i<j}\left[p_{ij}^{\alpha}(d_{ij})a_{ij}^{\alpha}+(1-p_{ij}^{\alpha}(d_{ij}))(1-a_{ij}^{\alpha})\right]
 \label{due}
 \eea
  where $p_{ij}^{\alpha}(d_{ij})$ is given by Eq. $(\ref{pex2})$, i.e.
 \bea
 p_{ij}^{\alpha}(d_{ij})=\frac{\theta_i^{\alpha} \theta_j^{\alpha} e^{-d/d_{\alpha}}}{1+ \theta_i^{\alpha} \theta_j^{\alpha} e^{-d/d_{\alpha}}}.
 \label{p2exp}
 \eea
The ``hidden variables" $\theta_i^{\alpha}$  fix the expected  degree of node $i$ in layer $\alpha$,
i.e.
\bea
\kappa_i^{\alpha}=\sum_j p_{ij}^{\alpha}(d_{ij}),\label{ka}
\eea
 while the ``hidden variables" $d_{\alpha}$  fix the total cost $L^{\alpha}=N\ell^{\alpha}$ associated with the links in layer $\alpha$ given by 
 \bea
 L^{\alpha}=N\ell^{\alpha}=\sum_{i<j}d_{ij}p_{ij}^{\alpha}(d_{ij}).
 \label{la}
 \eea

 In these multiplexes the expected total overlap $\Avg{O^{\alpha,\alpha'}}$ of the links between layer $\alpha$ and layer $\alpha'$ and the expected local overlap $\Avg{o_i^{\alpha,\alpha'}}$ of the links between layer $\alpha$ and layer $\alpha'$ are given by Eqs. $(\ref{AO})$, that we rewrite here for convenience,
 \bea
  \avg{O^{\alpha,\alpha'}}&=&\sum_{i<j}p^{\alpha}_{ij} p^{\alpha'}_{ij}\nonumber \\
 \avg{o_i^{\alpha,\alpha'}}&=&\sum_{j=1,j\neq i}^Np^{\alpha}_{ij} p^{\alpha'}_{ij}.
\label{AO2} 
 \eea
 Here we want to show that the expected total and local  overlap can be significant for the spatial multiplex ensemble under consideration.
 
 Let us for simplicity consider a multiplex in which the expected degrees in a certain layer  are all equal and finite. Moreover let us assume that the nodes are distributed uniformly on a $D$ dimensional Euclidean hypersphere of radius $R$,  with density $\rho$.  Therefore, we have $\kappa_i^{\alpha}=\kappa^{\alpha}\ \forall i$ and the so called ``hidden variables" in a given layer are the same for every node, i.e. $\theta_i^{\alpha}=\theta^{\alpha}\ \forall i$.  
 In this case we can easily estimate the relation between $(\kappa^{\alpha},L^{\alpha})$ and $(\theta^{\alpha},d_{\alpha})$.
 In fact approximating the sum over $j$ with an integral over a continuous distribution of points in Eq. $(\ref{ka})$, we find
 \bea
 \kappa^{\alpha}&\simeq &\rho \ \Omega(D)\int_{0}^{R}dr \ r^{D-1}\frac{\left(\theta^{\alpha}\right)^2e^{-r/d_{\alpha}}}{1+\left(\theta^{\alpha}\right)^2e^{-r/d_{\alpha}}}\nonumber\\
 &\simeq &\rho \ \Omega(D)\sum_{n=0}^{\infty} (-1)^n\left(\theta^{\alpha}\right)^{2(n+1)}\int _{0}^R dr \ r^{D-1} e^{-r (1+n)/d_{\alpha}}\nonumber \\
 &\simeq & \rho\  \Omega(D) \Gamma(D) d_{\alpha}^D\sum_{n=0}^{\infty} (-1)^n\frac{\left(\theta^{\alpha}\right)^{2(n+1)}}{(1+n)^D},
 \label{cal}
 \eea
 where $\Omega(D)r^{D-1}$ is the surface area of a $D$ dimensional hypersphere of radius $r$, and therefore $\Omega(D)$ is given by  $\Omega(D)=\frac{2\pi^{D/2}}{\Gamma\left(\frac{D}{2}\right)}$ and where we have assumed $\theta^{\alpha}e^{-r/d}<1$. Moreover in the large network limit we assume that $\Omega(D)R^D/D\simeq N$ and in the last expression of Eqs. $(\ref{cal})$ we have performed the limit $R\to \infty$. 
 The relation between $\kappa^{\alpha}$ and $(\theta^{\alpha},d_{\alpha})$ can be furthermore simplified as 
 \bea
 \kappa^{\alpha}\simeq-\rho\  \Omega(D)\ \Gamma(D)\  d_{\alpha}^D \mbox{Li}_{D}\left[-\left(\theta^{\alpha}\right)^2\right],
 \label{ka2}
 \eea
 where $Li_n(z)$ is the polylogarithmic function.
  Performing  similar calculations we can show that in the continuous approximation, where we approximate the sum on $(i,j)$ with an integral over space, we have that Eq.~$(\ref{la})$ can be written as
  \bea
  \frac{L^{\alpha}}{N}=\frac{1}{2}\ell^{\alpha}\simeq\rho \Omega(D+1)\Gamma(D+1)d_{\alpha}^{D+1} \mbox{Li}_{D+1}\left[-\left(\theta^{\alpha}\right)^2\right].
  \label{la2}\eea
 Since we are interested in the case in which both $\kappa^{\alpha}$ and $\ell^{\alpha}$ are finite, it follows from the Eqs.~$(\ref{ka2})-(\ref{la2})$, that the ``hidden variables" $(\theta^{\alpha}, d_{\alpha})$ are also finite, i.e. they do not depend on $N$ in the limit $N\to \infty$.
We can now easily evaluate the scaling with the total number of nodes $N$ of the expected total overlap between two layers $\Avg{O^{\alpha,\alpha'}}$ and the expected local overlap $\Avg{o^{\alpha,\alpha'}}$ between two layers using Eqs.~$(\ref{AO2})$.
In particular we have  in the continuous approximation, for the expected total overlap between layer $\alpha$ and layers $\alpha'$, 
\bea
\Avg{O^{\alpha,\alpha'}}&\simeq &N\frac{\rho}{2}\Omega(D)\int_0^R dr r^{D-1}\frac{\left(\theta^{\alpha}\right)^2e^{-r/d_{\alpha}}}{1+\left(\theta^{\alpha}\right)^2e^{-r/d_{\alpha}}}\times \nonumber\\
&&\times\frac{\left(\theta^{\alpha'}\right)^2e^{-r/d_{\alpha'}}}{1+\left(\theta^{\alpha'}\right)^2e^{-r/d_{\alpha'}}}.
\eea 
Performing straightforward calculations we get that 
\bea
\Avg{O^{\alpha,\alpha'}}\simeq N \frac{\rho}{2} \frac{\Omega(D)}{2} I(\alpha,\alpha')
\label{oexp1}
\eea
 where $I(\alpha,\alpha')$ is finite and in the limit $R,N\to \infty$ and is  given by 
 \bea
 I(\alpha,\alpha')&\simeq&\sum_{n=0}^{\infty}\sum_{m=0}^{\infty}(-1)^{m+n}\left(\theta^{\alpha}\right)^{2(n+1)}\left(\theta^{\alpha'}\right)^{2(m+1)}\times \nonumber \\
 &&\times\left(\frac{d_{\alpha}d_{\alpha'}}{d_{\alpha'}(1+n)+d_{\alpha}(1+m)}\right)^D.
 \label{Iaa'}
 \eea
 Therefore the expected total overlap between two layers is linear in $N$, i.e. a finite fraction of all the links is overlapping.
 Moreover it can be shown that the overlap is significant (finite) in every region of the network, as also the expected local overlap is significant.
 In fact, following similar steps used to estimate the expected total overlap we can show that 
 \bea
 \Avg{o^{\alpha,\alpha'}_i}\simeq \rho \Omega(D) I(\alpha,\alpha')
 \label{oexp2}
 \eea 
 with $I(\alpha,\alpha')$ given by Eq. $(\ref{Iaa'})$ in the limit $R,N\to \infty$. 
These results remain qualitatively the same if the multiplex is formed by networks with heterogeneous degree distribution.

\section{Interacting networks }
\subsection{Definition}
Interacting networks are formed by a set of networks of different nature and a set of links connecting nodes in different networks. An example of interacting networks is the airport network  and railway network in India, where airports and train stations are usually distinct, that we will study in detail in a subsequent section.
Therefore interacting networks are  a set of $M$ networks $G_{\alpha}=(V_{\alpha},E_{\alpha})$ with $\alpha=1,2\ldots, M$ where the set of nodes $V_{\alpha}$ is different for every network. 
In addition to this we have to consider also the interactions between the nodes in different networks. These interactions can be represented by a set  of bipartite networks such as  ${\cal G}_{\alpha,\beta}=(V_{\alpha} \cup V_{\beta},E_{\alpha,\beta})$ that connects the nodes of  a network $\alpha$ with  the nodes of another network $\beta$.
Therefore an ensemble of interacting networks will be given by
the set $(\vec{G},\vec{\cal G})=(\{G_{\alpha}\},\{{\cal G}_{\alpha,\beta}\})$.
In these types of networks we can have that one node $i$ in network $\alpha$ is linked to several nodes in network $\beta$, or that one node in network $\alpha$ is not linked to any node in network $\beta$. This feature of the network provides a further flexibility of these types of networks with respect to a multiplex where each node of the network is represented at the same time in different layers.
We note here that these types of networks are also very interesting to study diffusion processes, extending the work done for the  multiplex networks  in \cite{Diffusion}.
\subsection{Ensembles of spatial interacting networks}
The statistical mechanics treatment of spatial interacting networks follows closely the derivation of the spatial multiplex ensembles.
Spatial interacting networks ensembles  are ensembles of networks $(\vec{G},\vec{\cal G})=(\{G_{\alpha}\},\{{\cal G}_{\alpha,\beta}\})$.  Each network $G_{\alpha}=(V_{\alpha},E_{\alpha})$ with $\alpha=1,2\ldots, M$ is formed by a different set of  $N_{\alpha}$ nodes  embedded in a metric space. Each node  is assigned a coordinate $\vec{r}$ in this metric space. Each bipartite network ${\cal G}_{\alpha,\beta}$ connects nodes of network $\alpha$ with nodes of network $\beta$.  In general 
 a spatial  ensemble of interacting networks  is defined once we define the probability $P(\vec{G},\vec{\cal G}|\{\vec{r}\})$ of the interacting networks  $(\vec{G},\vec{\cal G})$ conditioned to the positions of the nodes $\{\vec{r}\}$.
For ensembles of spatial interacting networks the  {\it entropy} $S$  is given by 
\begin{equation}
S=-\sum_{\vec{G},\vec{\cal G}} P(\vec{G},\vec{\cal G}|\{\vec{r}\})\log P(\vec{G},\vec{\cal G}|\{\vec{r}\}).
\label{S4}
\end{equation}
Spatial interacting networks ensembles can be constructed by maximizing the 
entropy of the ensemble, while fixing a set  $K$ of soft constraints 
\begin{equation}
\sum_{\vec{G},\vec{\cal G}}F_{\mu}(\vec{G}, \vec{\cal G}|\{\vec{r}\})P(\vec{G},\vec{\cal G}|\{\vec{r}\})=C_{\mu},
\label{constr4}
\end{equation}
with $\mu=1,2\ldots, K$, and $F_{\mu}(\vec{G},\vec{\cal G}|\{\vec{r}\})$ being  a function of the multiplex and the positions of the nodes.
In this way it is easy to show that the probability $P(\vec{G},\vec{\cal G}|\{\vec{r}\})$ of a multiplex in this ensemble is given by 
\bea
P(\vec{G},\vec{\cal G}|\{\vec{r}\})=\frac{1}{Z}e^{-\sum_{\mu=1}^K \lambda_{\mu}F_{\mu}(\vec{G},\vec{\cal G}|\{\vec{r}\})},
\label{pg4}
\eea
where $Z$ is the normalization constant, and the values of the Lagrangian multipliers $\lambda_{\mu}$ for each constraint $\mu=1,2,\ldots, K$ are fixed by imposing the constraints in Eqs. $(\ref{constr4})$.
We consider here the special  case of spatial interacting networks ensembles  generated by this approach when each constraint $F_{\mu}(\vec{G},\vec{\cal G}|\{\vec{r}\})$ involve a single network. In this case $P(\vec{G},\vec{\cal G}|\{\vec{r}\})$ can be written as 
 \bea
 P(\vec{G},\vec{\cal G}|\{\vec{r}\})=\prod_{\alpha=1}^M P_{\alpha}(G_{\alpha}|\{\vec{r}\})\prod_{\alpha<\beta}P_{\alpha, \beta}({\cal G}_{\alpha, \beta}|\{\vec{r}\}).
 \label{sub}
 \eea
 where $P_{\alpha}(G_{\alpha}|\{\vec{r}\})$ is the probability of a network $G_{\alpha}$ in a maximal entropy ensemble and $P_{\alpha,\beta}({\cal G}_{\alpha,\beta}|\{\vec{r}\})$ is the probability of the bipartite network ${\cal G}_{\alpha,\beta}$ in a maximal entropy ensemble of a bipartite network.
 \begin{figure}
\centering 
\includegraphics[width=0.45\textwidth]{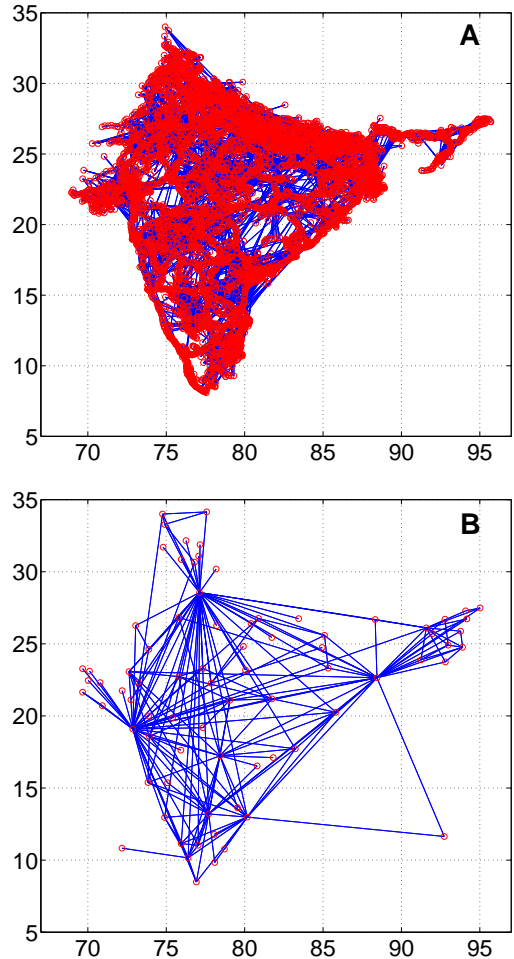}
\caption{Map of the Indian railway network (RR) (panel A) and map of the  Indian airport network (AA) (panel B).}
\label{maps}
\end{figure}
\subsection{The interacting airport  and railway   networks in India}

As a specific example of interacting networks we consider the air transportation network and the train transportation network in India. We have extracted the data of railway stations, train route and schedule of trains at different stations in the Indian railway \footnote{www.indianrail.gov.in}. Two stations are connected if there exists a physical track connecting the two stations, with the links corresponding to connections within one stop distance. There are $7408$ stations and $13230$ links in the railway network. The airport network is generated by drawing links between airports with direct flight connections between them. The data for flight schedule has been extracted from from the database of Indian airports \footnote{www.ourairports.com}. In our dataset we have $78$ airports with $203$ links. 
Additionally we access the data of the bipartite network of interconnections between airports and train stations from the website $indianrailinfo.com$. Here we have accessed only those airports which are commercially used for passenger travel and we have extracted the information about a railway station and a nearby airport. The information about a nearby airport is provided if there exists a road access between the train station and the airport. There are $6769$ rail stations and $102$ airports mentioned in the database of interconnections between airports and train stations, out of which $78$ airports are commercially used. We therefore drop the remaining $24$ airports from our analysis. Additionally we have accessed the latitude and longitude of the airports and of the railway stations using Google maps (see Figure $\ref{maps}$ displaying the maps of the railway network and the airport network under consideration).

\begin{figure}
\centering 
\includegraphics[width=0.4\textwidth]{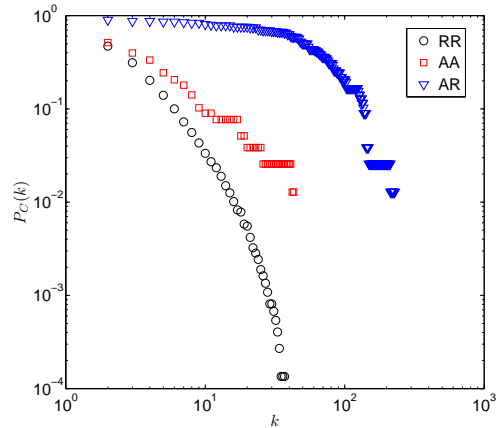}
\caption{Cumulative degree distribution of the Indian railway network (RR), the Indian airport network (AA) and the cumulative degree distribution of the airports in the AR bipartite network between airports and railway stations.}
\label{Pck}
\end{figure}

Therefore the set of interacting networks is formed by  the India airport network (the AA network), by  the India railway network (the  RR network) and by the bipartite network of interconnections between airports and train stations (the AR network). 
The cumulative degree distributions of the railway network (RR), the airport network (AA) and the airport degree distribution in the AR networks are shown in Figure $\ref{Pck}$. We note that the AA network is broad while the degree distribution of the railway network (RR) is not broad. Interestingly enough, the degree distribution of the airports in the AR network is also broad. We note that the degrees of the railway stations in the AR networks are either one or zero leading to a trivial degree distribution.
\begin{figure}
\centering 
\includegraphics[width=0.4\textwidth]{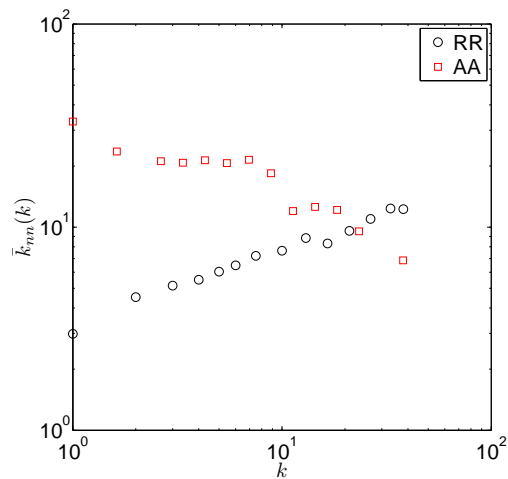}
\caption{The average nearest neighbor degree ${k}_{nn}(k)$ defined in Eq.~$(\ref{defknn})$  for the Indian airport network (AA) and the Indian railway network (RR). While the RR network is assortative, the AA network is disassortative.}
\label{knn}
\end{figure}

The degree correlations in the two interacting networks AA and RR are very different. In order to show this, we  plot in Figure $\ref{knn}$ the  function $k_{nn}(k)$ also called the average degree of the neighbor of a node of degree $k$,  defined as
\bea
k_{nn}(k)=\frac{1}{NP(k)}\sum_{i|k_i=k}\sum_{j}a_{ij}^{\alpha}k_j,
\label{defknn}
\eea
for network AA ($\alpha=1$) and for network RR ($\alpha=2$). While the railway network RR is assortative, and characterized by an increasing function $k_{nn}(k)$ the airport network AA is disassortative and characterized by a decreasing function $k_{nn}(k)$. Therefore highly connected airports tend to be linked to low connectivity airports, while highly connected railway stations are more likely to be connected to highly connected railway stations.
Moreover, in order to characterize other types of correlations, we  measure the Pearson coefficient $\rho$ between the degree $k^{AA}$ of an airport in the AA network and the degree $k^{AR}$ of the same airport in the AR network, i.e.
\bea
\rho=\frac{\Avg{k^{AA}k^{AR}}-\Avg{k^{AA}}\Avg{k^{AR}}}{\sqrt{\Avg{(k^{AA})^2}-\Avg{k^{AA}}^2}\sqrt{\Avg{(k^{AR})^2}-\Avg{k^{AR}}^2}}.
\eea
The calculated Pearson coefficient is  $\rho=0.3998$ indicating that the degree of the airports in the AA network is correlated with the degree of the airports in the AR network, enhancing the importance and centrality of high degree airports in this set of  interacting networks.

\begin{figure}
\centering 
\includegraphics[width=0.4\textwidth]{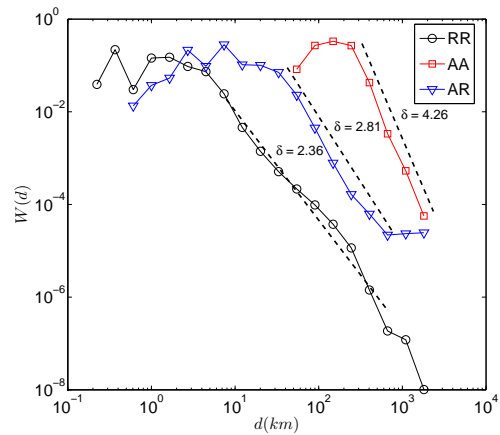}
\caption{Plot of $W(d)$ the  factor that depends on the distance $d$ between the nodes and that affects the link
probabilities, for the Indian railway network (RR), Indian airport network (AA) and the bipartite network of interconnections between airports and train stations (AR) . At large distance the functions $W(d)$ for the three networks decay as a power-law of distance $W(d)\propto d^{-\delta}$, with the value of the fitted exponent $\delta$ indicated in the figure.  }
\label{Wddelta}
\end{figure}

Finally we  consider the ensemble of interacting networks with $M=2$ in which $P(G_1,G_2,{\cal G}_{12})=P_1(G_1|\{\vec{r}\})P_2(G_2|\{\vec{r}\})P_{12}({\cal G}_{12}|\{\vec{r}\})$. The probabilities $P_{1,2}(G_{1,2}|\{\vec{r}\})$ are the probabilities of spatial networks in which the expected degree of each node is equal to the one observed respectively in the AA network and in the RR network and in which the expected average number of links at  a  given distance is equal to the one observed respectively in the AA network and in the RR network. The probability $P_{12}({\cal G}_{12}|\{\vec{r}\})$ is the probability of a bipartite network in the  ensemble of bipartite networks in which the expected degree of every node  is equal to the one observed in the AR network and in which the expected number of links at a given distance is  equal to the one observed in the AR network. 
In particular the link probabilities within each layer are given by Eq.~$(\ref{pW})$ and the link probabilities in the bipartite network are given by Eq~$(\ref{pWb})$.  In Figure $\ref{Wddelta}$ we show the functions $W(d)$ derived in Eq. $(\ref{pW})-(\ref{pWb})$, which depends on the distance between the nodes and affects link probabilities, for the networks AA, RR and AR.
We show that the function $W(d)$ at large distances decays as a power-law $W(d)\propto d^{-\delta}$ for the three cases under consideration and we indicate the fitted values of the exponents $\delta$  in the Figure $\ref{Wddelta}$. This shows that all these networks allow for long-range connections and  therefore the entire interacting network displays a good  navigability. We notice that the airport  network  (the AA network) is characterized by a $\delta$ exponent roughly twice as big as the railway network (RR network). However,  the airport network doesn't have any links at distances smaller than $10^2$ kilometers, while the maximal distance in this  dataset is limited because we consider only  connections within India,  therefore the probability of a long range airport connection is still larger than the probability of a long distance train connection.

\section{Conclusions}
In this paper we have introduced the statistical mechanics of spatial multiplex ensembles and of spatial interacting networks ensembles. This approach can be used to characterize a large variety of multiplexes and interacting networks embedded either in a real or in a hidden space. We have shown that spatial multiplexes, unlike uncorrelated sparse multiplexes, naturally develop a significant overlap of the links. Therefore the empirical observations of significant overlap occurring in multiplex datasets, such as in transportation multiplexes and social multiplexes, can be  caused by their underlying geometry.
Finally we have built ensembles of spatial interacting networks, and we have characterized an example of such structures: the interacting railway and airport  networks in India.
In the framework of the theory of randomized spatial interacting network ensembles, we have measured  the function $W(d)$ that modulates the link probability between two nodes at distance $d$ in the randomized airport (AA), railway (RR) and AR networks, showing  that the function $W(d)$ decays as a power-law of distance for large distances  in all the cases (AA, RR, AR networks).

Our analysis could be extended to directed and weighted networks. For example, the railway and air transportation networks could be generalized to weighted networks where the weight of each link is given by the number of 
trains (or flights) between two nodes. Complex spatial multiplex networks and spatial interacting networks are usually  co-evolving and inter-dependent  as it is demonstrated in the case of  a well integrated transportation system where the transfer from railway stations  to airports and vice versa should be efficient.
Earlier studies have dealt with onset of interdependence  in Chinese and European railway-airline transportation networks \cite{darenhe}.  However it lacked the spatial feature between the layers of inter dependent networks. In the future, we plan to extend our analysis by developing a generalized model to predict efficient functioning of various multiplex networks. 

In conclusion, we believe that modelling spatial multiplexes and spatial interacting networks will be essential for the investigation of  major complex systems as the brain, infrastructures, and social networks that cannot be fully understood if we do not characterize  their complex multilayer structure.

\appendix

\section{Expected overlap in multiplex ensembles with link probability decaying  like a power-law  with distance}
In order to generalize the results proven in paragraph $\ref{par}$ , we  evaluate here the expected overlap  for   a multiplex ensemble with link probability decaying as a power-law of the distance. In particular the link probability in the generic layer $\alpha$ satisfies Eq.~$(\ref{su})$,  where each $P_{\alpha}(G_{\alpha}|\{\vec{r}_i\})$ is given by 
 Eq. $(\ref{due})$ and 
  where $p_{ij}^{\alpha}(d_{ij})$ is given by Eq. $(\ref{ppow2})$, i.e.
 \bea
 p_{ij}^{\alpha}(d_{ij})=\frac{{\theta_i^{\alpha}\theta_j^{\alpha}} r^{-\delta_{\alpha}}}{1+ {\theta_i^{\alpha}\theta_j^{\alpha}} r^{-\delta_{\alpha}}}.
 \label{p2pow}
 \eea
The ``hidden variables" $\theta_i^{\alpha}$  fix the expected  degree of node $i$ in layer $\alpha$,
i.e.
\bea
\kappa_i^{\alpha}=\sum_j p_{ij}^{\alpha}(d_{ij}),
\eea
 and the ``hidden variables" $\delta_{\alpha}$  fix the total cost $L^{\alpha}=N\ell^{\alpha}$ given by Eq. $(\ref{c2})$ associated with the links in layer $\alpha$ that we rewrite here for convenience
 \bea
 L^{\alpha}=N\ell^{\alpha}=\sum_{i<j}\log(d_{ij})p_{ij}^{\alpha}(d_{ij}).
 \label{lapow}
 \eea
 
 Let us consider for simplicity the case in which all the expected degrees in the same layer are equal and finite, i.e. $\kappa_i^{\alpha}=\kappa^{\alpha}\ \  \forall i$. Moreover let us make the additional assumption that the nodes are distributed uniformly in a $D$ dimensional Euclidean hypersphere of radius $R$,  with density $\rho$.
In this hypothesis, following a procedure similar to the one presented in detail in paragraph $\ref{par}$, we get that  the relation between  $(\kappa^{\alpha},L^{\alpha}=N\ell^{\alpha})$ and the``hidden variables" $(\theta^{\alpha},\delta_{\alpha})$ is given, in the continuous approximation and in the limit $R,N\to \infty$ by 
\bea
\kappa^{\alpha}&=&\rho \Omega(D) \sum_{n=0}^{\infty}\left(\theta^{\alpha}\right)^{2(1+n)}\frac{(-1)^n}{\delta_{\alpha}(1+n)-D}\label{k3} \\
\ell^{\alpha}&=&\rho \frac{\Omega(D)}{2} \sum_{n=0}^{\infty}\left(\theta^{\alpha}\right)^{2(1+n)}\frac{(-1)^n}{\left[\delta_{\alpha}(1+n)-D\right]^2}\label{la3}
\eea
as long as $\delta_{\alpha}>D$.
Therefore, the ''hidden variables" $(\theta^{\alpha},\delta_{\alpha})$ are finite.
The expected total and local overlap between layer $\alpha $ and layer $\alpha'$ are given  by Eqs.($\ref{AO}$) that we can estimate in the continuous approximation and in the thermodynamic limit $R,N\to \infty$.
We have in particular
\bea
\Avg{O^{\alpha,\alpha'}}&=&N \frac{1}{2}\rho \Omega(D) J(\alpha,\alpha')\nonumber\\
\Avg{o^{\alpha,\alpha'}}&=&\rho \Omega(D) J(\alpha,\alpha')
\label{opow}
\eea
where $J(\alpha,\alpha')$ is finite and given by 
\bea
J(\alpha,\alpha') &=& \sum_{n=0}^{\infty}\sum_{m=0}^{\infty}\left(\theta^{\alpha}\right)^{2(n+1)}\left(\theta^{\alpha'}\right)^{2(m+1)}\times \nonumber \\
 &&\times\frac{1}{\delta_{\alpha}(1+n)+\delta_{\alpha'}(1+m)-D}.\nonumber \\
 \label{J}
\eea
Given the  Eqs.~$(\ref{opow})$ we can conclude that also in this case a finite fraction of links are overlapping between any two layers and that this overlap is distributed uniformly over the network.
\section{Expected overlap in multiplexes with some networks with link probability decaying exponentially with distance  and with the other networks with link probability decaying as a power-law}
Here we  evaluate the expected overlap in  multiplex ensembles with some networks with link probability decaying exponentially with distance  and with the other networks with link probability decaying as a power-law of the distance between the linked nodes. In particular the different layers  will have a link probability satisfying Eq.~$(\ref{su})$,  where the probabilities $P_{\alpha}(G_{\alpha}|\{\vec{r}_i\})$ are given by 
 Eq.~$(\ref{due})$ 
  where $p_{ij}^{\alpha}(d_{ij})$ for some layers is given by  Eq.~$(\ref{pex2})$, for other layers is given Eq.~$(\ref{ppow2})$.
  In other words the link probability in some layers is decaying exponentially with distance and in some other layers is decaying as a power-law of the distance. The ``hidden variables"  $\theta_i^{\alpha}$  fix the expected  degree of node $i$ in layer $\alpha$,
i.e.
\bea
\kappa_i^{\alpha}=\sum_j p_{ij}^{\alpha}(d_{ij}),
\eea
 and the ``hidden variables" $\delta_{\alpha}$ or $d_{\alpha}$  fix the total cost $L^{\alpha}=N\ell^{\alpha}$  associated with the links in layer $\alpha$ given by 
 \bea
 L^{\alpha}=N\ell^{\alpha}=\sum_{i<j}f_{\alpha}(d_{ij})p_{ij}^{\alpha}(d_{ij}).
 \label{lapow2}
 \eea
 where $f_{\alpha}(d_{ij})=\log(d_{ij})$ or $f_{\alpha}=d_{ij}$ depending on the layer $\alpha$.
 Let us consider for simplicity the case in which all the expected degrees in the same layer are equal and finite, i.e. $\kappa_i^{\alpha}=\kappa^{\alpha}\ \  \forall i$. Moreover let us make the additional assumption that the nodes are distributed uniformly in a Euclidean $D$ dimensional hypersphere of radius $R$,  with density $\rho$. For each network in each layer the ``hidden variables" $(\theta^{\alpha},d_{\alpha})$ can be found using the  Eqs.~$(\ref{ka2})-(\ref{la2})$, while the ``hidden variables" $(\theta^{\alpha},\delta_{\alpha})$ can be found using the Eqs.~$(\ref{k3}),(\ref{la3})$. If we consider two layers  with link probability decaying exponentially with distance we have that their expected global and local overlap is given by Eqs~$(\ref{oexp1})-(\ref{oexp2})$, if we have two layers  with link probability decaying as a power-law we find instead Eqs.~$(\ref{opow}),(\ref{J})$. Finally if we have two layers, a layer $\alpha$ with link probability decaying exponentially with distance, and a layer $\alpha'$ with link probability decaying as a power-law of the distance between the nodes, the expected  total and global overlap between these two layers is given by 
\bea
\Avg{O^{\alpha,\alpha'}}&=&N \frac{\rho}{2} \Omega(D) K(\alpha,\alpha')\nonumber\\
\Avg{o^{\alpha,\alpha'}}&=&\rho \Omega(D) K(\alpha,\alpha')
\label{opowexp}
\eea
where $K(\alpha,\alpha')$ is finite and given by 
\bea
K(\alpha,\alpha')&=& \sum_{n=0}^{\infty}\sum_{m=0}^{\infty}\left(\theta^{\alpha}\right)^{2(n+1)}\left(\theta^{\alpha'}\right)^{2(m+1)}\times \nonumber \\
 &&\times E_{1+\delta_{\alpha'} (1+m)-D}\left(\frac{1}{d_{\alpha}}\right),\nonumber \\
\eea
where $E_n(z)$ is the exponential integral function.
Therefore, also in the case in which a spatial multiplex is formed by  some networks with link probability decaying exponentially with the distance and other networks with link probability decaying as a power-law of the distance, the expected global and local overlap is significant.

\end{document}